\documentclass[3p,twocolumn,amsmath,amssymb]{elsarticle}
\usepackage{graphicx}
\usepackage{epsfig}
\def\lsim{\raise0.3ex\hbox{$\;<$\kern-0.75em\raise-1.1ex\hbox{$\sim\;$}}}
\def\gsim{\raise0.3ex\hbox{$\;>$\kern-0.75em\raise-1.1ex\hbox{$\sim\;$}}}
\newcommand{\be}{\begin{eqnarray}}
\newcommand{\ee}{\end{eqnarray}}

\def\bea{\begin{eqnarray}}
\def\eea{\end{eqnarray}}

\begin{document}


\title{On the top-antitop invariant mass spectrum at the LHC\\[0.15cm]
from a Higgs boson signal perspective}

%

\author{S. Moretti and D.A. Ross}
\address{{\it School of Physics and Astronomy, University of Southampton,
Highfield, Southampton SO17 1BJ, UK.}}

\date{\today}

\begin{abstract}
We investigate the effect of one-loop corrections of 
${\cal O}(\alpha_S^2\alpha_W)$ on the $t\bar t$ invariant mass spectrum at the Large
Hadron Collider (LHC)
in presence of both resonant and non-resonant Higgs boson effects. We show that 
corrections of ${\cal O}(\alpha_S^2\alpha_W)$ involving a non-resonant Higgs boson
are comparable to or even larger than those involving interference with
the $s$-channel resonant Higgs boson amplitude
and that both of these are subleading with respect to all other (non-Higgs) diagrams through that
order. We also compute the contribution through ${\cal O}(\alpha_S^2\alpha_W^2)$ of resonant
Higgs boson production (i.e. Higgs production via $gg$ fusion) as well as the pure QCD ones
of ${\cal O}(\alpha_S^3)$. Altogether, we show that
the well known peak-dip structure of the $M_{t\bar t}$ spectrum emerging from interference effects
between the $t,u$-channel $gg$-induced Leading Order (LO) QCD diagrams and the one due to a Higgs boson  
in $s$-channel via $gg$-fusion is drastically swamped by the remainder of the terms 
of ${\cal O}(\alpha_S^2\alpha_W)$ discussed above. 
\end{abstract}

\maketitle
\section{Introduction}
\noindent
Higgs decay to top-antitop pairs could be a search channel in a variety of models where the corresponding
Branching Ratio (BR) becomes dominant, owing to the fact that,
 in some such models, the competing Higgs decays into $W^+W^-$
and $ZZ$ pairs can be suppressed or non-existent. For example, in the case of the Minimal Supersymmetric
Standard Model (MSSM), this naturally occurs if the decaying Higgs boson is the CP-odd one, $A$, but also for the 
heavy CP-even one, $H$, when $\cos(\beta-\alpha)\approx0$ (see \cite{HHG}
for the definition of $\alpha$ and $\beta$ and the relevant MSSM Feynman rules). In this case, which is generally realised for 
$M_A\approx M_H\gsim 200$ GeV, $A,H\to t\bar t$ becomes a feasible search channel \cite{ATLAS-paper}. 
Unfortunately, LEP (and now also Tevatron and LHC) limits almost
exclude entirely the region of the $(M_A,\tan\beta)$ plane of the MSSM where significances above 5 can be 
achieved \cite{ATLAS-TDR,CMS-TDR}. 
On the other hand, in a generic 2-Higgs Doublet Model (2HDM), one can easily switch 
off the 
Higgs couplings to gauge boson pairs, as the trigonometric factor
which controls this coupling
 is essentially a free 
parameter (unlike the case of the MSSM), whilst still
 being consistent with LEP (and more recent Tevatron
and LHC) data \cite{HHG}. Furthermore,
 even in those models where the scope of the Higgs $\to t\bar t$ decay channel may 
be limited, once a Higgs boson is found and its parameters (mass primarily, but possibly also width and discrete quantum numbers) measured, these can subsequently be exploited in the selection of the  $ t\bar t$ final state
 to render the Higgs visible \cite{profile}. This can be done,
for example, for
the purpose of measuring the corresponding BR,
thereby accessing the Higgs-$t\bar t$ coupling, in the attempt to profile the
new resonance. 

Therefore, it remains of the utmost importance to continue assessing the scope
of existing colliders in extracting the Higgs $\to t\bar t$ signal, not least because, 
from a theoretical point of view, the Yukawa 
coupling between a Higgs boson and the top quark entering its rates is a parameter which generally
 plays a key role in the dynamics 
of any Beyond the Standard Model (BSM) scenario based on a Higgs mechanism: e.g., in its
 Electro-Weak Symmetry 
Breaking (EWSB) transition, in the Renormalisation Group Equations (RGEs) describing its evolution from high to 
low energy scales, etc.  Finally, one should recall that the 
top(anti-top)-quark decays into a 
bottom(anti-bottom)-quark and a $W^\pm$ boson rather than hadronising,
thereby transmitting its spin properties to the
decay products very efficiently, so that the spin and CP nature
of the Higgs boson(s) involved  can be explored 
in suitable experimental observables \cite{ttpol,review}. 

Clearly, in order to perform all the relevant searches  
for Higgs $\to t\bar t$ events, any source of SM corrections
should be well-understood. This is true for the case of the Higgs signal,
$gg\to {\rm Higgs}\to t\bar t$ \cite{PMZ,ggEW}. Regarding the QCD background, $q\bar q,gg\to t\bar t$, 
complete one-loop results exist for both the QCD 
\cite{QCD-SM-tt} (see also \cite{earlycalc-QCD}) and EW \cite{EW-SM-tt} (see also \cite{earlycalc-EW}) 
sectors. However, it is also of crucial importance to take into account interference effects between the
above signal and background (limited to the $gg$ channel) \cite{Higgs-paperi}. 

Concerning such an interference, it is well known in the literature \cite{Higgs-paperi} that Higgs
 bosons produced via $gg$ 
fusion and decaying into
top-antitop quark pairs interfere  with  the background  $t,u$-channel LO QCD diagrams to
produce a peak-dip structure in the $ t\bar t$ invariant mass spectrum. This 
structure has been claimed as potentially observable at the LHC.
We point out here that, if one properly includes through the perturbative order at which this effect
emerges, i.e., ${\cal O}(\alpha_S^2\alpha_W)$, all other diagrams, such an effect is generally subleading
with respect to those induced by all other (non-resonant) Higgs graphs appearing through that order 
and/or the ones due to all other (non-Higgs) diagrams, the latter being always the dominant contribution.
(We neglect here  the $q\bar q\to t\bar t $ channel, as this is negligible at the LHC and
does not involve a resonant Higgs contribution.)

The plan of the paper is as follows. In the next section we describe our calculation. In Sect. III
we report on our numerical results. Finally, we conclude in Sect. IV.  

\begin{figure}
\includegraphics[width=0.95\linewidth]{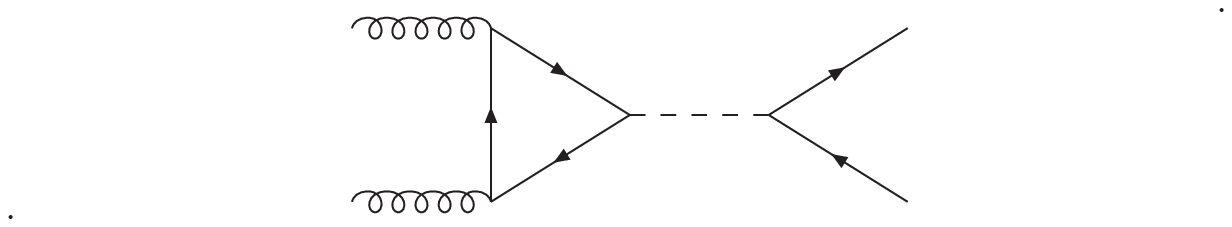}
\caption{$gg \, \to \, t \bar{t}$ via a Higgs produced in the $s$-channel.}
\label{fdh1} \end{figure}

\begin{figure}
\includegraphics[width=0.95\linewidth]{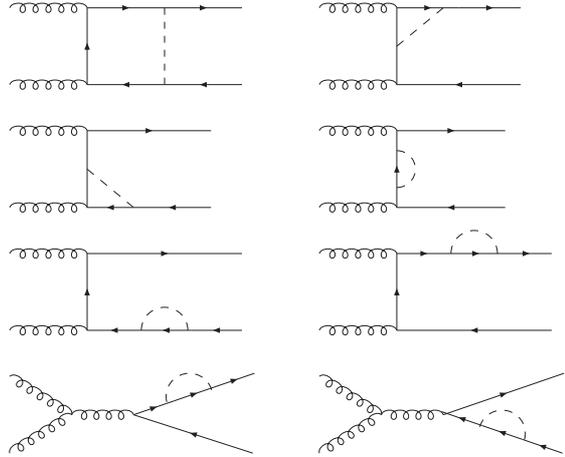}
\caption{Remaining graphs involving a Higgs in the process
$gg \, \to \, t \bar{t}$ (excluding gluon crossings).}
\label{fdh2} \end{figure}

%
\section{Calculation}
\noindent
The computation exploited in this paper is described at length in Ref.~\cite{ourselves},
so we do not dwell here on the technical details. Rather, we just recall the salient features of 
the calculation,
with a view at emphasising the role of the diagrams with an intermediate Higgs boson.

For the complete EW corrections we need to consider all graphs
containing a weakly (or electromagnetically) interacting internal particle.
Graphs involving an internal Higgs are shown in Figs.~\ref{fdh1}--\ref{fdh2},
where Fig.~\ref{fdh1} is the graph for the Higgs produced in the $s$-channel.
This can interfere with the LO QCD amplitude ($t$- and $u$-channel exchanges only
since interference with $s$-channel exchange is forbidden by colour).
Fig. \ref{fdh2} shows all the other graphs involving a Higgs, which can also
interfere with (the complete) LO QCD amplitude.
We want to compare  the yield of these two interferences as well
as to that of the entire set of EW corrections
(the remaining graphs have the same topology as the graphs in Figs. \ref{fdh1}--\ref{fdh2},
but with the Higgs replaced by a gauge and/or Goldstone boson.)
The rates
emerging through Next-to-LO (NLO) QCD will also be presented for reference.
Notice that we only include here weak effects, thereby neglecting one-loop diagrams
(and accompanying tree-level bremsstrahlung graphs) involving a photon. These were computed 
in \cite{EW-SM-tt} and found to be negligibly small.
This is expected since QED corrections (after accounting for bremsstrahlung) 
 see no large enhancement at the LHC, unlike the case of the weak ones,
which are affected by large so-called Sudakov logarithms of ${\cal O}(\log^2({\hat{s}}/M_W^2))$. 

For the top mass and width
 we have taken $m_t=175$ GeV and $\Gamma_t=1.55$ GeV, respectively. 
The $Z$ mass used was $M_Z=91.19$ GeV and was related to the $W$ mass, $M_W$, via the
SM formula $M_W=M_Z\cos\theta_W$, where $\sin^2\theta_W=0.232$.
(Corresponding widths were $\Gamma_Z=2.5$ GeV and $\Gamma_W=2.08$ GeV.)
The Parton Distribution Functions (PDFs) that we have used are the CTEQ6 set
\cite{cteq6} taken at the factorisation/renormalisation scale $Q=\mu=2m_t$. (We also have checked
other sets, but found no significant difference in the relative size of our
corrections.) The choice of PDFs dictates the running and parameters used
to compute $\alpha_S$, which  has been taken at one-loop  level in the calculation
of terms of ${\cal O}(\alpha_S^2)$ and
${\cal O}(\alpha_S^2\alpha_W)$ (in conjunction with CTEQ6L1)
 and  at two-loop level in the computation of effects of ${\cal O}(\alpha_S^3)$
(in conjunction with CTEQ6M). The value used throughout for the electromagnetic
coupling was $\alpha=1/128$ (with $\alpha_W=\alpha/\sin\theta_W$). 
The NLO QCD corrections have been included through 
MCFM \cite{MCFM}. Finally, the Higgs decay widths have been computed by means of the programs
described in Refs.~\cite{Moretti:1994ds} and \cite{Kunszt:1996yp}. 
All rates are presented at the LHC energy of 14 TeV, though
the qualitative effects seen here would not change  at 7 or at 8 TeV.

\section{Numerical results}
\subsection{CP-even Higgs state}
\begin{figure}
 \centering\begin{tabular}{c}
  \includegraphics[scale=0.35,angle=90]{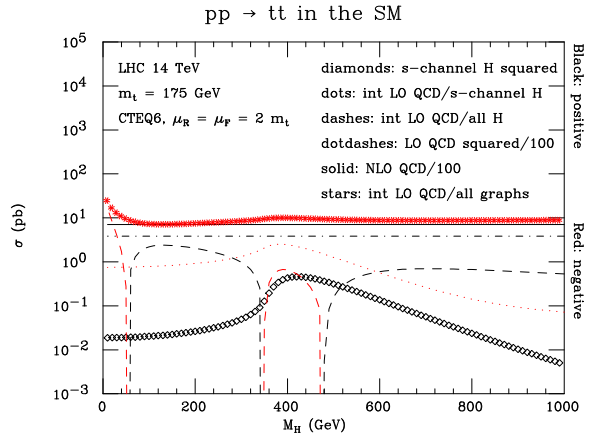}
 \end{tabular}
\caption{The integrated cross section at the LHC for all the components of the $gg\to t\bar t$ process
described in the text. The Higgs processes are for the CP-even state of the SM. Where the curves are red it signifies that they
correspond to a negative contribution (notice that the $y$-axis is logarithmic).}
\label{fig:sigmaH}
\end{figure}

As the  benchmark for the CP-even Higgs boson case we take the SM. Fig. \ref{fig:sigmaH} shows 
the contributions to the inclusive cross section
from various different perturbative components.
 Clearly, the pure QCD contribution through ${\cal O}(\alpha_S^3)$ is
largest (solid line). (In the plots it also incorporates the lowest order ${\cal O}(\alpha_S^2)$ term,
dot-dashed line.)
The ${\cal O}(\alpha_S^2\alpha_W)$ interference between the $t,u$-channel LO QCD diagrams with the
one-loop $s$-channel Higgs graph is the dotted curve, which is always negative and largest for
$M_H$ just above $2m_t$. However, the dashed line, which refers to the interference with {\it all}
graphs involving a Higgs boson, indicates that this term is actually positive and dominated
by the graphs other than the $s$-channel Higgs exchange except in the region of $M_H$
between 250 and 550 GeV, where the interference is indeed negative but
has an absolute value  at least an order of
magnitude smaller than that obtained from the $s$-channel Higgs exchange only.
 While this is a potential welcome feature through the 
${\cal O}(\alpha_S^2\alpha_W)$, as it would increase the total signal yield,
over a considerable range of $M_H$  (notice that
the square of the $s$-channel Higgs graph through ${\cal O}(\alpha_S^2\alpha_W)$, diamond line,
is consistently one order of magnitude or so smaller than its interference with the LO QCD
terms), it becomes negligible if one includes all other non-Higgs diagrams
through ${\cal O}(\alpha_S^2\alpha_W)$. In fact, the complete result at such a perturbative
level, the starred line, is always negative and significantly larger than the partial
correction obtained by only including all Higgs diagrams.   
    
\begin{figure}
 \centering\begin{tabular}{c}
  \includegraphics[scale=0.35,angle=90]{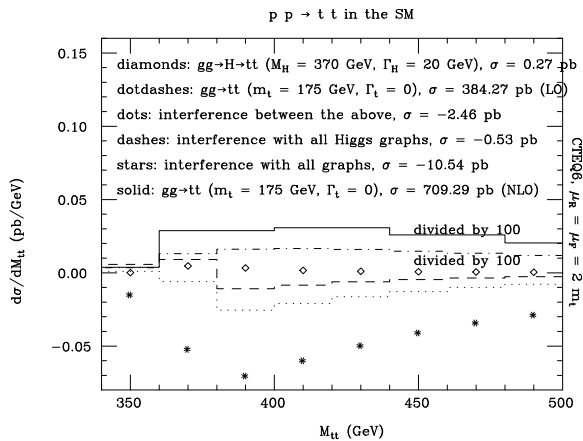}
 \end{tabular}
\caption{The differential cross section in invariant mass
at the LHC for all the components of the $gg\to t\bar t$ process
described in the text. The Higgs processes are for the CP-even state of the SM
and assume $M_H=370$ GeV. Bin width is 20 GeV.}
\label{fig:MassH370}

\end{figure}

\begin{figure}
 \centering\begin{tabular}{c}
  \includegraphics[scale=0.35,angle=90]{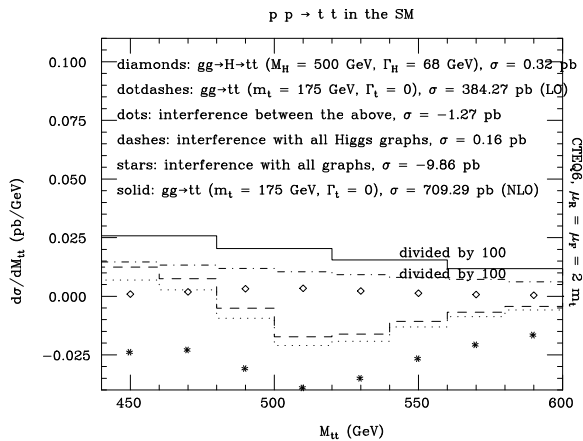}
 \end{tabular}
\caption{Same as Fig. \ref{fig:MassH370} for $M_H=500$ GeV.}
\label{fig:MassH500}
\end{figure}

Figs. \ref{fig:MassH370} and \ref{fig:MassH500} show the differential distributions in $t\bar t$ invariant mass for two 
Higgs mass values, one just above the on-shell $t\bar t$ threshold ($M_H=370$ GeV)
and another well above it ($M_H=500$ GeV), respectively. The overall trend here is the same as the
one just established at inclusive level, i.e., the salient feature is again that the
full ${\cal O}(\alpha_S^2\alpha_W)$ effects swamp those involving the $s$-channel
Higgs diagram only. Curiously though, the aforementioned peak-dip structure is maintained
(compare the starred histogram with the dotted one), albeit a factor approximately 5(8) larger
(and still negative) for $M_H=370(500)$ GeV. Therefore, the ultimate effect 
through the complete ${\cal O}(\alpha_S^2\alpha_W)$ is to swamp the Higgs resonance
emerging through  ${\cal O}(\alpha_S^2\alpha_W^2)$. 
(Notice that the cross sections in the legends 
of Figs. \ref{fig:MassH370}--\ref{fig:MassH500} are the integrated ones over the entire invariant mass range.)  

\subsection{CP-odd Higgs state}

\begin{figure}
 \centering\begin{tabular}{c}
  \includegraphics[scale=0.35,angle=90]{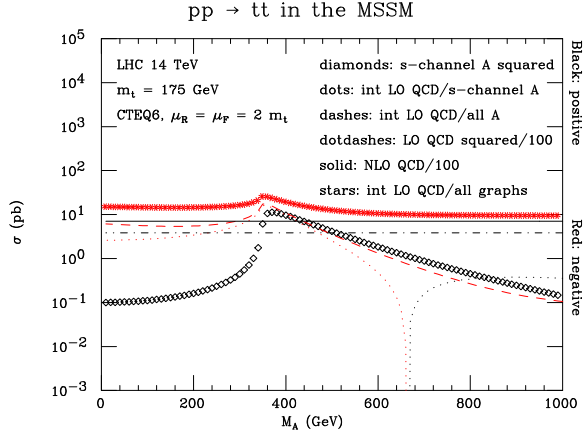}
 \end{tabular}
\caption{The integrated cross section at the LHC for all the components of the $gg\to t\bar t$ process
described in the text. The Higgs processes are for the CP-odd state of the MSSM. Where the curves are red it signifies that they
correspond to a negative contribution (notice that $y$-axis is logarithmic).}
\label{fig:sigmaA}
\end{figure}

As the benchmark for the CP-odd Higgs boson case we take the MSSM with
$\tan\beta=1$ and Supersymmetric matter decoupled. Notice in this case the much reduced width
of the Higgs boson, with respect to the SM setup, owing to the absence of $A\to W^+W^-$ and $ZZ$ decays. 
Fig.~\ref{fig:sigmaA} shows the inclusive cross section in its usual components. Here, since the 
BR$(A\to t\bar t)$ is much
increased compared to the SM case, the $s$-channel Higgs diagram dominates the contributions
from other graphs involving a Higgs boson
 (compare the dotted and dashed curves). Nevertheless,
the complete ${\cal O}(\alpha_S^2\alpha_W)$ result is far larger than the terms involving Higgs
mediation only, much on the same footing as in the SM. 

\begin{figure}
 \centering\begin{tabular}{c}
  \includegraphics[scale=0.35,angle=90]{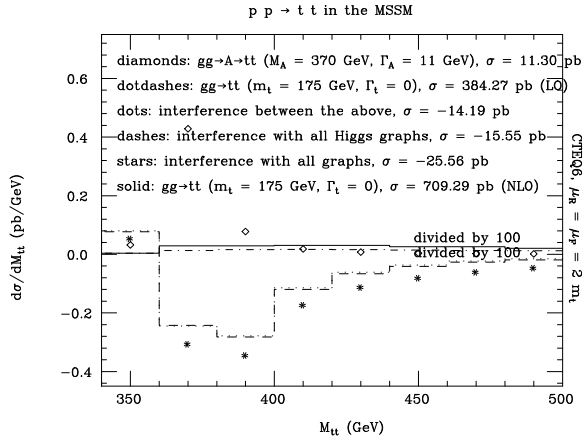}
 \end{tabular}
\caption{The differential cross section in invariant mass
at the LHC for all the components of the $gg\to t\bar t$ process
described in the text. The Higgs processes are for the CP-odd state of the MSSM
and assume $M_A=370$ GeV. Bin width is 20 GeV.}
\label{fig:MassA370}
\end{figure}

\begin{figure}
\vspace*{0.5cm}
 \centering\begin{tabular}{c}
  \includegraphics[scale=0.35,angle=90]{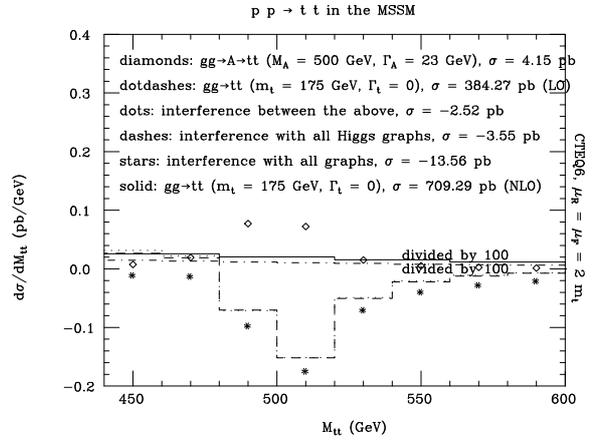}
 \end{tabular}
\caption{Same as Fig. \ref{fig:MassA370} for $M_A=500$ GeV.}
\label{fig:MassA500}
\end{figure}

Figs. \ref{fig:MassA370} and \ref{fig:MassA500} display the usual $M_{t\bar t}$ spectra
for $M_A=370$ and 500 GeV, respectively. Again, also for this differential cross section, 
we see  the leading role of the $s$-channel Higgs diagram in the ${\cal O}(\alpha_S^2\alpha_W)$ 
terms, as it accounts for more than 90\% of the Higgs contributions for $M_A=370$ GeV (compare dotted and dashed
histograms, respectively). However, again in this case,
 the remainder of the loop diagrams entering the ${\cal O}(\alpha_S^2\alpha_W)$
result is larger. Furthermore, all these terms have always the same (negative) sign, so that they strongly contribute to deplete
the yield of the ${\cal O}(\alpha_S^2\alpha_W^2)$ signal (compare star and diamond symbols). The situation
at $M_A=500$ GeV is not only qualitatively similar but also quantitatively, as the $s$-channel Higgs contribution is 
more than 70\% of the total due to Higgs graphs up to $M_A \approx 670$ GeV, after which the interference with
the $s$-channel Higgs changes sign. 
The overall result is that
the full ${\cal O}(\alpha_S^2\alpha_W)$ result is again much bigger (a factor of more than 4) than the one due to the
$s$-channel Higgs diagram alone. Altogether, again, on sees a strong depletion
 of the ${\cal O}(\alpha_S^2\alpha_W^2)$ 
Higgs signal once the full ${\cal O}(\alpha_S^2\alpha_W)$ result is accounted for.    

\section{Conclusions and outlook}
We have shown that interference effects between LO QCD amplitudes 
and one-loop diagrams involving weak interactions 
for the $gg\to t\bar t$ process can be large at the LHC and in fact overwhelming the
well studied interference between the former and the component of the latter due to 
the $s$-channel resonant Higgs diagram only. The overall effect is generally to diminish the significance
of  the
Higgs signal produced as a resonance in the  gluon-gluon channel. Therefore,
these previously overlooked contributions
need to be included in realistic simulations aiming at establishing the LHC potential to
extract a Higgs $\to t\bar t$ signature. We have illustrated this at the 14 TeV stage of the LHC
using both a CP-even and CP-odd Higgs boson, the former belonging to the SM and the latter to
the MSSM. For further analyses, we make our program available upon request.

\section*{Acknowledgments}
SM is partially supported through the NExT Institute.
DAR thanks the Theory Unit at CERN for its hospitality during the preparation of this manuscript.

\end{document}